# ON THE ASYMPTOTIC REPRESENTATION FOR TRANSVERSE MAGNETIC MULTIPLE SCATTERING OF RADIATION BY AN INFINITE GRATING OF DIELECTRIC CIRCULAR CYLINDERS AT OBLIQUE INCIDENCE

ÖMER KAVAKLIOĞLU and BARUCH SCHNEIDER
*Division of Electrophysics Research and Department of Mathematics,*
*Faculty of Computer Sciences, Izmir University of Economics,*
*Balçova, IZMIR 35330 TURKEY*
(omer_kavaklioglu@yahoo.com; omer.kavaklioglu@ieu.edu.tr; baruch.schneider@ieu.edu.tr)

**Abstract**

In this article, we present the derivation of the *'asymptotic forms'* of the equations corresponding to the *'scattering coefficients of the exterior electric and magnetic fields of an infinite grating of insulating dielectric circular cylinders for vertically polarized and obliquely incident plane electromagnetic waves'*. Exploiting the generalized forms of the *"Twersky's elementary function representations for Schlömilch series"*, we have deducted an *'Ansatz'* describing the behavior of the scattering coefficients of the electric and magnetic fields for obliquely incident waves when the grating spacing is much smaller than the wavelength of the incident electromagnetic radiation. Introducing the statement of this *'Ansatz'* into the equations of the *'scattering coefficients of the infinite grating at oblique incidence'*, and expanding the scattering coefficients in the form of an *'asymptotic series'* as a function of the ratio of the radius of the cylinders to the grating spacing, we have acquired two *'new'* infinite sets of algebraic equations associated with the *'scattering coefficients of the exterior electric and magnetic fields of the grating for vertically polarized and obliquely incident plane waves'*.

---

## 1. Introduction

Lord Rayleigh (1881) first treated the classical electromagnetic problem of the incidence of plane electric waves on an insulating dielectric cylinder as long ago as 1881. He published the diffraction of a plane wave at normal incidence by a homogeneous dielectric cylinder (Lord Rayleigh 1881). His solution was generalized for obliquely incident plane waves when the magnetic vector of the incident wave is transverse to the axis of the cylinder by Wait (1955).

Twersky (1952a) first obtained the formal analytical solution for the scattering of a plane electromagnetic wave by an arbitrary configuration of parallel cylinders in terms of cylindrical wave functions, considering all possible contributions to the excitation of a particular cylinder by the radiation scattered by the remaining cylinders in the grating. In his solution, he expressed the scattered wave as an infinite sum of orders of scattering, and later extended his solution to consider the case where all the axes of cylinders lie in the same plane (Twersky 1952b). Twersky (1952c) then introduced the "multiple scattering theories" to the finite grating of cylinders, and employed "Green's function methods" to represent the "multiple scattering amplitude of one cylinder within the grating" in terms of "the functional equation" and the "single scattering amplitude of an isolated cylinder" (Twersky 1956). Twersky (1962) obtained a set of algebraic equations for the multiple scattering coefficients in terms of the elementary function representations of Schlömilch series (Twersky 1961), and in terms of the known coefficients of an isolated cylinder.

Bogdanov *et al.* (1985a) constructed an algorithm for the problem of diffraction of a plane electromagnetic wave, incident arbitrarily on a periodic array of infinitely long dielectric rods of circular cross section, and presented the relations between the main diffraction characteristics of the array and its parameters. Bogdanov *et al.* (1985b, 1987, 1991) treated various configurations of the same problem.

More recent investigations in the area of scattering by the arrays of cylinders have been conducted by Nicorovici *et al.* (1994) who developed the spatial and spectral domain forms of the Green's function for the diffraction of a plane wave at arbitrary incidence in the x-y plane on a grating oriented along the x axis. Nicorovici and McPhedran (1994) considered the spatial and spectral domain forms of the Green's function appropriate in the electromagnetic diffraction of a plane wave incident at an arbitrary angle in the x-y plane on a singly periodic structure oriented along the x-axis, and established the expressions from which grating lattice sums can effectively be evaluated. In addition, Chin *et al.* (1994) investigated the techniques for representing in absolutely convergent forms of the lattice sums in doubly periodic electromagnetic diffraction problems. Petit (1980) presented a more generalized case of arbitrary incidence and discussed quasi-periodicity.

Problems dealing with two dimensional arrays have been treated in detail by McPhedran *et al.* (2000) who investigated the lattice sums arising in quasi-periodic Green's functions, McPhedran and Nicorovici (2002) who investigated sums arising in doubly quasiperiodic Green's functions, McPhedran *et al.* (2004) who studied two-dimensional lattice sums, McPhedran *et al.* (2005a) who considered sums over the square lattice and provide formulas, McPhedran *et al.* (2005b) who took into consideration the use of Poisson summation formula to obtain effective formulas for sums arising in scattering problems for the case of an infinite number of cylinders ordered periodically along a line in the form of an infinite array.

These theoretical ideas mentioned above have found substantial applications in the studies of Botten *et al.* (2000) who developed a formulation for wave propagation and scattering through stacked gratings comprising metallic and dielectric cylinders. Furthermore, Botten *et al.* (2004) developed a semi-analytic approach for analyzing photonic crystals by employing the Bloch mode scattering matrix methods and White *et al.* (2004) applied this method to two-dimensional photonic waveguide structures that consist of lattices of either parallel finite dielectric cylinders in an air background or parallel finite air cylinders submerged in a dielectric medium.

Cai and Williams (1999a, b) investigated the multiple scattering of anti-plane shear waves in fiber-reinforced composite materials, and Cai (2006) treated the *'layered multiple scattering method'* for anti-plane shear wave scattering from multiple gratings consisting of parallel cylinders.

Previous investigations mentioned above do not include the most general case of oblique incidence although the grating is illuminated by an incident electromagnetic wave at an arbitrary angle to the x-axis. As far as can be ascertained by the writers, Sivov (1961) first treated the diffraction by an infinite periodic array of perfectly conducting cylindrical columns for the most generalized case of obliquely incident plane polarized electromagnetic waves in order to determine the reflection and transmission coefficients of the infinite grating of perfectly conducting cylinders in free space. The period of the grating was assumed to be small in comparison with the wavelength. Lee (1990) studied the scattering of an obliquely incident electromagnetic wave by an arbitrary configuration of parallel, non-overlapping infinite cylinders and presented the solution for the scattering of an obliquely incident plane wave by a collection of closely-spaced, radially-stratified parallel cylinders that can have an arbitrary number of stratified layers (Lee 1992). Kavaklıoğlu (2000, 2001, 2002) and Kavaklıoğlu and Schneider (2007) extended the results of Twersky (1956, 1962) for the multiple scattering of an obliquely incident plane electromagnetic wave by an infinite grating of dielectric circular cylinders. In a more recent investigation by Kavaklıoğlu (2007), the *'direct Neumann iteration technique'* is employed in order to acquire the exact solutions for the scattering coefficients of an infinite grating in the form of an infinite series and an analogue of Twersky's solution is acquired for obliquely incident plane electromagnetic waves.

The most generalized oblique incidence solution presented in this investigation, the direction of the incident plane wave makes an arbitrary oblique angle of arrival $\theta_i$ with the positive *z*-axis as indicated in figure 1.

## 2. Problem formulation

2.1. *Multiple scattering representations for an infinite grating of dielectric circular cylinders for obliquely incident E-polarized plane electromagnetic waves*

A vertically polarized plane electromagnetic wave, which is obliquely incident upon the infinite array of insulating dielectric circular cylinders having infinite length with radius "$a$", dielectric constant "$\varepsilon_r$", and relative permeability "$\mu_r$", can be expanded (Wait 1955; kavaklıoğlu 2000) in the cylindrical coordinate system $(R_s, \phi_s, z)$ of the $s^{th}$ cylinder in terms of the cylindrical waves referred to the axis of $s^{th}$ cylinder as

$$\mathbf{E}_v^{inc}(R_s,\phi_s,z) = \hat{\mathbf{v}}_i E_{0v} e^{ik_r sd \sin\psi_i} \left\{ \sum_{n=-\infty}^{\infty} e^{-in\psi_i} J_n(k_r R_s) e^{in(\phi_s + \pi/2)} \right\} e^{-ik_z z} \tag{1}$$

The cylinders of the grating are placed perpendicularly to the *x*-*y* plane, and separated by a distance of "$d$", as indicated in figure 1. In the above description of the incident field, $\hat{\mathbf{v}}_i$ denotes the vertical polarization vector associated with a unit vector having a component parallel to all the cylinders, $\phi_i$ is the angle of incidence in *x*-*y* plane measured from $x-$axis in such a way that $\psi_i = \pi + \phi_i$, implying that the wave is obliquely incident in the first quadrant of the coordinate system, and "$J_n(x)$" stands for "Bessel function of order *n*." In addition, we have the following definitions

$$k_r = k_0 \sin\theta_i \tag{2a}$$

$$k_z = k_0 \cos\theta_i \tag{2b}$$

"$e^{-i\omega t}$" time dependence is suppressed throughout the paper, where "$\omega$" stands for the angular frequency of the incident wave in radians per second and "*t*" represents time in seconds.

*2.2. Expressions for the z-components of the exterior fields*

The centers of the cylinders in the infinite grating are located at positions $\mathbf{r}_0$, $\mathbf{r}_1$, $\mathbf{r}_2$,.., etc. The exact solution for the *z*-components of the electric field in the exterior of the grating belonging to this configuration can be expressed in terms of the incident electric field in the coordinate system of the $s^{th}$ cylinder located at $\mathbf{r}_s$, plus a summation of cylindrical waves outgoing from each of the individual $m^{th}$ cylinder located at $\mathbf{r}_m$, as $|\mathbf{r}-\mathbf{r}_m| \to \infty$, i. e.,

$$E_z^{(ext)}(R_s,\phi_s,z) = E_z^{inc}(R_s,\phi_s,z) + \sum_{m=-\infty}^{+\infty} E_z^{(m)}(R_m,\phi_m,z) \tag{3}$$

The external electric and magnetic field intensities associated with vertically polarized obliquely incident plane electromagnetic waves are then given in (Kavaklıoğlu 2000) as

$$E_z^{(ext)}(R_s,\phi_s,z) = \Bigg\{ e^{ik_r sd \sin\psi_i} \sum_{n=-\infty}^{+\infty} \Bigg[ \Bigg( E_n^i + \sum_{m=-\infty}^{\infty} A_m \, I_{n-m}(k_r d) \Bigg) J_n(k_r R_s)$$

$$+A_n H_n^{(1)}(k_r R_s) \Bigg] e^{in(\phi_s+\pi/2)} \Bigg\} e^{-ik_z z} \qquad (4a)$$

$$H_z^{(ext)}(R_s,\phi_s,z)=\Bigg\{ e^{ik_r sd\sin\psi_i} \sum_{n=-\infty}^{+\infty} \Bigg[ \Bigg( \sum_{m=-\infty}^{\infty} A_m^H \, \mathit{l}_{n-m}(k_r d) \Bigg) J_n(k_r R_s)$$

$$+A_n^H H_n^{(1)}(k_r R_s) \Bigg] e^{in(\phi_s+\pi/2)} \Bigg\} e^{-ik_z z} \qquad (4b)$$

In the representation of the electric and magnetic fields above, $\{A_n, A_n^H\}_{n=-\infty}^{\infty}$ denotes the set of all multiple scattering coefficients of the infinite grating associated with *"vertically polarized obliquely incident plane electromagnetic waves"*, $\forall n \ni n \in Z$, where *"Z"* represents the set of all integers. In expressions (4a, b), we have

$$E_n^i = \sin\theta_i E_{0v} e^{-in\psi_i} \qquad (5a)$$

$$\mathit{l}_n(2\pi\Delta)=\sum_{p=1}^{+\infty} H_n^{(1)}(2\pi p\Delta)\left[e^{2\pi i p\Delta\sin\psi_i}(-1)^n + e^{-2\pi i p\Delta\sin\psi_i}\right] \qquad (5b)$$

where $\Delta \equiv \frac{k_r d}{2\pi}$, and "$H_n^{(1)}(x)$" denotes the $n^{th}$ order Hankel function of first kind, $\forall n \ni n \in Z$, where "**Z**" represents the set of all integers. The series in expression (5b) is the generalized form of the *'Schlömilch series for obliquely incident waves* $\mathit{l}_{n-m}(k_r d)$' (Twersky 1961, Kavaklıoğlu 2002) and convergent provided that $k_r d(1\pm\sin\psi_i)/2\pi$ does not equal integers.

## 3. Derivation of the Asymptotic Equations for the Scattering Coefficients of the Infinite Grating at Oblique Incidence

This section is devoted to the derivation of the asymptotic equations for the scattering coefficients of the infinite grating of dielectric cylinders at oblique incidence. In order to demonstrate the procedure of obtaining the asymptotic equations, we have introduced the exact system of equations for the scattering coefficients $\{A_n;\ A_n^H\}_{n=-\infty}^{+\infty}$ of the infinite grating of dielectric circular cylinder associated with an obliquely incident vertically polarized plane wave by the application of the boundary conditions on the surface of each cylinder within the grating in Kavaklıoğlu (2000) as

$$b_n^{\mu}\left\{ A_n + c_n\left[ E_n^i + \sum_{m=-\infty}^{+\infty} A_m \, I_{n-m}(k_r d) \right] \right\} = -\left[ A_n^H + a_n^{\mu} \sum_{m=-\infty}^{+\infty} A_m^H \, I_{n-m}(k_r d) \right] \tag{6a}$$

$\forall n \ni n \in Z$ , and

$$b_n^{\varepsilon}\left[ A_n^H + c_n \sum_{m=-\infty}^{+\infty} A_m^H \, I_{n-m}(k_r d) \right] = A_n + a_n^{\varepsilon}\left[ E_n^i + \sum_{m=-\infty}^{+\infty} A_m \, I_{n-m}(k_r d) \right] \tag{6b}$$

$\forall n \ni n \in Z$ . The coefficients arising in this infinite set of linear algebraic equations are defined as

$$c_n := \frac{J_n(k_r a)}{H_n^{(1)}(k_r a)} \tag{7}$$

$\forall n \ni n \in Z$ , and two sets of constants $a_n^{\zeta}$ and $b_n^{\zeta}$ , in which $\zeta_r \in \{\varepsilon_r, \mu_r\}$ stands for the relative permittivity and permeability of the dielectric cylinders respectively , are given as

$$a_n^{\zeta} = \left[ \frac{J_n(k_1 a)\dot{J}_n(k_r a) - \zeta_r\left(\frac{k_r}{k_1}\right) J_n(k_r a)\dot{J}_n(k_1 a)}{J_n(k_1 a)\dot{H}_n^{(1)}(k_r a) - \zeta_r\left(\frac{k_r}{k_1}\right) H_n^{(1)}(k_r a)\dot{J}_n(k_1 a)} \right] \tag{8}$$

for $\zeta \in \{\varepsilon, \mu\}$ , and $\forall n \ni n \in Z$ ; where $k_1$ is defined as $k_1 = k_0\sqrt{\varepsilon_r \mu_r - \cos^2\theta_i}$ , and

$$b_n^{\zeta} = \sqrt{\frac{\varepsilon_0 \mu_0}{\zeta_0^2}} \left[ \frac{J_n(k_1 a) H_n^{(1)}(k_r a)}{J_n(k_1 a)\dot{H}_n^{(1)}(k_r a) - \zeta_r\left(\frac{k_r}{k_1}\right) H_n^{(1)}(k_r a)\dot{J}_n(k_1 a)} \right] \left(\frac{inF}{k_r a}\right) \tag{9}$$

for $\zeta \in \{\varepsilon, \mu\}$, and $\forall n \ni n \in Z$, where $F$ in the expression above is a constant and given as

$$F = \frac{(\mu_r \varepsilon_r - 1)\cos\theta_i}{\mu_r \varepsilon_r - \cos^2\theta_i} \tag{10}$$

$\forall n \ni n \in Z$. In these equations $\varepsilon_r$ and $\mu_r$ denotes the dielectric constant and the relative permeability constant of the insulating dielectric cylinders; $\varepsilon_0$ and $\mu_0$ stands for the permittivity and permeability of the free space respectively, $A_n$ and $A_n^H$ correspond to the scattering coefficients for the electric field intensity and magnetic field intensity associated with obliquely incident plane E-polarized electromagnetic waves, respectively. The $J_n'$, and $H_n'^{(1)}$ in expressions (7-8) are defined as

$$J_n'(\varsigma) \equiv \tfrac{d}{d\varsigma} J_n(\varsigma) \tag{11a}$$

$$H_n'^{(1)}(\varsigma) \equiv \tfrac{d}{d\varsigma} H_n^{(1)}(\varsigma) \tag{11b}$$

which imply the first derivatives of the Bessel and Hankel functions of first kind and of order *n* with respect to their arguments.

### *3. 1. Derivation of the Approximate Equations for the Scattering Coefficients of the Infinite Grating at Oblique Incidence*

The exact equations in (6a-b) can be solved for $A_n$ and $A_n^H$ when the distance between the cylinders of the infinite grating are smaller than the wavelength of the incident wave, i. e., for $k_r d << 1$ the exact equations take the following form

$$\begin{pmatrix} A_{\pm n} \\ A_{\pm n}^H \end{pmatrix} \cong \underline{\underline{S_n}} \begin{pmatrix} E_{\pm n}^i + \sum_{m=-\infty}^{\infty} A_m H_{\pm n-m}(k_r d) \\ \sum_{m=-\infty}^{\infty} A_m^H H_{\pm n-m}(k_r d) \end{pmatrix} \tag{12}$$

where $\underline{\underline{S_n}}$ is a $(2\times 2)$ matrix defined as

$$\underline{\underline{S}}_n := \begin{pmatrix} s_n^{\varepsilon\mu} & s_{\pm n}^{\xi} \\ s_{\pm n}^{\eta} & s_n^{\mu\varepsilon} \end{pmatrix} \frac{(k_r a)^{2n}}{D} \tag{13}$$

and '$H_n(k_r d)$' stands for the approximations to the *"exact form of the Schlömilch series $I_n(k_r d)$"* in the limiting case when for $k_r d << 1$. Introducing (13) into (12), *"the approximate set of equations for the scattering coefficients of the infinite grating at oblique incidence"* can explicitly be written as

$$\begin{pmatrix} A_{\pm n} \\ A_{\pm n}^{H} \end{pmatrix} \cong \frac{(k_r a)^{2n}}{D} \begin{pmatrix} s_n^{\varepsilon\mu} & s_{\pm n}^{\xi} \\ s_{\pm n}^{\eta} & s_n^{\mu\varepsilon} \end{pmatrix} \begin{pmatrix} E_{\pm n}^{i} + \sum_{m=-\infty}^{\infty} A_m H_{\pm n-m}(k_r d) \\ \sum_{m=-\infty}^{\infty} A_m^{H} H_{\pm n-m}(k_r d) \end{pmatrix} \tag{14}$$

In the above, we have

$$D = \left[1 + \varepsilon_r \left(\frac{k_r}{k_1}\right)^2\right]\left[1 + \mu_r \left(\frac{k_r}{k_1}\right)^2\right] - F^2 \tag{15}$$

The $n$-dependent constants appearing in (13-14) are defined as

$$s_n^{\varepsilon\mu} := \left[\frac{in\pi}{(2^n n!)^2}\right] s_{\varepsilon\mu} \tag{16a}$$

$$s_{\pm n}^{\xi} := \left[\frac{in\pi}{(2^n n!)^2}\right] s_{\pm\xi} \tag{16b}$$

$$s_{\pm n}^{\eta} := \left[\frac{in\pi}{(2^n n!)^2}\right] s_{\pm\eta} \tag{16c}$$

$$s_n^{\mu\varepsilon} := \left[\frac{in\pi}{(2^n n!)^2}\right] s_{\mu\varepsilon} \tag{16d}$$

$\forall n \ni n \in N$ , where $N$ denotes the set of all natural numbers. The various constants appearing in the definitions (16) are expressed as

$$s_{\varepsilon\mu} = \left[1 - \varepsilon_r \left(\frac{k_r}{k_1}\right)^2\right]\left[1 + \mu_r \left(\frac{k_r}{k_1}\right)^2\right] + F^2 \tag{17a}$$

$$s_{\mu\varepsilon} = \left[1 - \mu_r \left(\frac{k_r}{k_1}\right)^2\right]\left[1 + \varepsilon_r \left(\frac{k_r}{k_1}\right)^2\right] + F^2 \tag{17b}$$

and

$$s_{\pm\xi} = \pm 2i\xi_0 F \tag{18a}$$

$$s_{\pm\eta} = \mp 2i\eta_0 F \tag{18b}$$

The elements of the matrix of coefficients in (14) can be calculated using the expressions (17-18), for instance $\left(\frac{s_{\varepsilon\mu}}{D}\right)$ and $\left(\frac{s_{\mu\varepsilon}}{D}\right)$ terms can be written as

$$\frac{s_{\varepsilon\mu}}{D} \equiv \frac{\left[1-\varepsilon_r\left(\frac{\sin^2\theta_i}{\mu_r\varepsilon_r-\cos^2\theta_i}\right)\right]\left[1+\mu_r\left(\frac{\sin^2\theta_i}{\mu_r\varepsilon_r-\cos^2\theta_i}\right)\right]+\left[\frac{(\mu_r\varepsilon_r-1)\cos\theta_i}{\mu_r\varepsilon_r-\cos^2\theta_i}\right]^2}{\left[1+\varepsilon_r\left(\frac{\sin^2\theta_i}{\mu_r\varepsilon_r-\cos^2\theta_i}\right)\right]\left[1+\mu_r\left(\frac{\sin^2\theta_i}{\mu_r\varepsilon_r-\cos^2\theta_i}\right)\right]-\left[\frac{(\mu_r\varepsilon_r-1)\cos\theta_i}{\mu_r\varepsilon_r-\cos^2\theta_i}\right]^2} \tag{19a}$$

$$\frac{s_{\mu\varepsilon}}{D} \equiv \frac{\left[1-\mu_r\left(\frac{\sin^2\theta_i}{\mu_r\varepsilon_r-\cos^2\theta_i}\right)\right]\left[1+\varepsilon_r\left(\frac{\sin^2\theta_i}{\mu_r\varepsilon_r-\cos^2\theta_i}\right)\right]+\left[\frac{(\mu_r\varepsilon_r-1)\cos\theta_i}{\mu_r\varepsilon_r-\cos^2\theta_i}\right]^2}{\left[1+\varepsilon_r\left(\frac{\sin^2\theta_i}{\mu_r\varepsilon_r-\cos^2\theta_i}\right)\right]\left[1+\mu_r\left(\frac{\sin^2\theta_i}{\mu_r\varepsilon_r-\cos^2\theta_i}\right)\right]-\left[\frac{(\mu_r\varepsilon_r-1)\cos\theta_i}{\mu_r\varepsilon_r-\cos^2\theta_i}\right]^2} \tag{19b}$$

In terms of the definitions of (16), *"the approximate set of equations for the scattering coefficients of the infinite grating at oblique incidence"* given in (14) takes the following form

$$\begin{pmatrix} A_{\pm n} \\ A^H_{\pm n} \end{pmatrix} \cong \frac{1}{D}\begin{pmatrix} s_{\varepsilon\mu} & s_{\pm\xi} \\ s_{\pm\eta} & s_{\mu\varepsilon} \end{pmatrix}\begin{pmatrix} E^i_{\pm n} + \sum\limits_{m=-\infty}^{\infty} A_m H_{\pm n-m}(k_r d) \\ \sum\limits_{m=-\infty}^{\infty} A^H_m H_{\pm n-m}(k_r d) \end{pmatrix}\left[\frac{in\pi}{(2^n n!)^2}\right](k_r a)^{2n} \tag{20}$$

$\forall n \ni n \in N$. These equations can be separated into two different sets, in which the first one contains only the odd coefficients and the second one contains only the even coefficients, as

$$A_{\pm(2n-1)} \cong \frac{(k_r a)^{4n-2}}{D}[s^{\varepsilon\mu}_{2n-1}(E^i_{\pm(2n-1)} + \sum_{m=-\infty}^{\infty} \mathrm{H}_{\pm(2n-1)-m} A_m)$$
$$+ s^{\xi}_{\pm(2n-1)}(\sum_{m=-\infty}^{\infty} \mathrm{H}_{\pm(2n-1)-m} A^H_m)] \tag{21a}$$

$$A^H_{\pm(2n-1)} \cong \frac{(k_r a)^{4n-2}}{D}[s^{\eta}_{\pm(2n-1)}(E^i_{\pm(2n-1)} + \sum_{m=-\infty}^{\infty} \mathrm{H}_{\pm(2n-1)-m} A_m)$$
$$+ s^{\mu\varepsilon}_{2n-1}(\sum_{m=-\infty}^{\infty} \mathrm{H}_{\pm(2n-1)-m} A^H_m)] \tag{21b}$$

$\forall n \ni n \in N$ for the odd coefficients. Similarly, for the even coefficients, we have acquired the following two sets of infinite number of approximate equations for the undetermined scattering coefficients of the infinite grating as

$$A_{\pm 2n} \cong \frac{(k_r a)^{4n}}{D}[s_{2n}^{\varepsilon\mu}(E_{\pm 2n}^{i} + \sum_{m=-\infty}^{\infty} \mathsf{H}_{\pm 2n-m} A_m) + s_{\pm 2n}^{\xi}(\sum_{m=-\infty}^{\infty} \mathsf{H}_{\pm 2n-m} A_m^H)] \tag{22a}$$

$$A_{\pm 2n}^{H} \cong \frac{(k_r a)^{4n}}{D}[s_{\pm 2n}^{\eta}(E_{\pm 2n}^{i} + \sum_{m=-\infty}^{\infty} \mathsf{H}_{\pm 2n-m} A_m) + s_{2n}^{\mu\varepsilon}(\sum_{m=-\infty}^{\infty} \mathsf{H}_{\pm 2n-m} A_m^H)] \tag{22b}$$

$\forall n \ni n \in N$.

### *3. 2. Derivation of the approximate expressions of the 'Schlömilch series $\mathsf{H}_n = \mathsf{J}_n + i\,\mathsf{N}_n$' in the limit of $k_r d << 1$*

The elementary function representations of the *'Schlömilch series* $\mathsf{I}_n(k_r d)$*'* given in (5b) have been originally derived by Twersky (1956) for the *'normal incidence',* and modified by Kavaklıoğlu (2002) for the *'oblique incidence'*. We will employ these elementary function representations for the evaluation of the asymptotic forms of the *'Schlömilch series* $\mathsf{H}_n = \mathsf{J}_n + i\,\mathsf{N}_n$*'* in the limit of $k_r d << 1$.

Twersky's forms (Twersky 1956) are still valid for the case of obliquely incident waves (Kavaklıoğlu 2002) with a slight modification in their arguments. We have obtained $\mathsf{H}_0$ for the special case of $n = 0$ as

$$\mathsf{H}_0 = -1 + \frac{1}{\pi\Delta}\sum_{\mu=-\mu_-}^{\mu_+}\frac{1}{\cos\phi_\mu} + \frac{2}{i\pi}\ln\frac{\Delta\gamma}{2} + \frac{i}{\pi}\left(\sum_{\mu=1}^{\mu_+} + \sum_{\mu=1}^{\mu_-}\right)\frac{1}{\mu} + \frac{1}{i\pi}\sum_{\mu=\mu_++1}^{\infty}\left[\frac{1}{\Delta\sinh\eta_\mu^+} - \frac{1}{\mu}\right] + \frac{1}{i\pi}\sum_{\mu=\mu_-+1}^{\infty}\left[\frac{1}{\Delta\sinh\eta_\mu^-} - \frac{1}{\mu}\right] \tag{23}$$

and for the general case, we have derived $\mathsf{H}_n$, $\forall n \ni n \in N$ as

$$\mathsf{H}_{2n} = \frac{1}{\pi\Delta}\sum_{\mu=-\mu_-}^{\mu_+}\frac{\cos 2n\phi_\mu}{\cos\phi_\mu} + \frac{i}{\pi}\left[\frac{1}{n} + \sum_{m=1}^{n}\frac{(-1)^m 2^{2m}(n+m-1)!}{(2m)!(n-m)!}\frac{B_{2m}(\Delta\sin\psi_i)}{\Delta^{2m}}\right] + \frac{1}{i\pi\Delta}\left[\left(\sum_{\mu=0}^{\mu_+} - \sum_{\mu=-1}^{-\mu_-}\right)\frac{\sin 2n\phi_\mu}{\cos\phi_\mu} + (-1)^n\left(\sum_{\mu=\mu_++1}^{\infty}\frac{e^{-2n\eta_\mu^+}}{\sinh\eta_\mu^+} + \sum_{\mu=\mu_-+1}^{\infty}\frac{e^{-2n\eta_\mu^-}}{\sinh\eta_\mu^-}\right)\right] \tag{24a}$$

$$\mathsf{H}_{2n+1} = \frac{1}{i\pi\Delta}\sum_{\mu=-\mu_-}^{\mu_+}\frac{\sin(2n+1)\phi_\mu}{\cos\phi_\mu} + \frac{2}{\pi}\sum_{m=0}^{n}\frac{(-1)^m 2^{2m}(n+m)!}{(2m+1)!(n-m)!}\frac{B_{2m+1}(\Delta\sin\psi_i)}{\Delta^{2m+1}} \tag{24b}$$

$$+\frac{1}{\pi\Delta}\left[\left(\sum_{\mu=0}^{\mu_+}-\sum_{\mu=-1}^{-\mu_-}\right)\frac{\cos(2n+1)\phi_\mu}{\cos\phi_\mu}+(-1)^{n+1}\left(\sum_{\mu=\mu_++1}^{\infty}\frac{e^{-(2n+1)\eta_\mu^+}}{\sinh\eta_\mu^+}-\sum_{\mu=\mu_-+1}^{\infty}\frac{e^{-(2n+1)\eta_\mu^-}}{\sinh\eta_\mu^-}\right)\right]$$

where $\mu_\pm$ is the upper and lower bounds for the propagating modes, and $\eta_\mu^\pm$'s are determined from the grating equation as $\cosh\eta_\mu^\pm = \pm\sin\psi_i + \frac{\mu}{\Delta}$.

*3. 2. 1. Approximations for* $\mathsf{H}_0$ , $\mathsf{H}_{2n}$ *and* $\mathsf{H}_{2n+1}$ *in the limit of* $\Delta << 1$

The real part of $H_n$, $\forall n \ni n \in \mathbf{Z}_+$ in (23, 224a, b), which is recognized as *'Bessel series* $J_n$*'*, can explicitly be written as

$$J_{2n} = \left[\frac{2}{k_r d}\sum_{\mu=-\mu_-}^{\mu_+}\frac{\cos 2n\phi_\mu}{k_r d\cos\phi_\mu} - \delta_{n0}\right]; \quad \forall n \ni n \in \mathbf{Z}_+, \tag{25a}$$

$$J_{2n+1} = \left[\frac{2}{ik_r d}\sum_{\mu=-\mu_-}^{\mu_+}\frac{\sin(2n+1)\phi_\mu}{\cos\phi_\mu}\right]; \quad \forall n \ni n \in \mathbf{Z}_+, \tag{25b}$$

where $\mathbf{Z}_+ = \{0, 1, 2, 3, \ldots.\}$.

*3. 2. 2. Approximations for* $\mathsf{N}_0$ , $\mathsf{N}_{2n}$ *and* $\mathsf{N}_{2n+1}$ *in the limit of* $\Delta << 1$

The imaginary part of the *'Schlömilch series* $H_n$*'*, which is known as the *'Neumann series* $N_n$*'* in (3), can be put into the following form for this limiting case as

$$\mathsf{N}_0 \cong -\frac{2}{\pi}\ln\frac{\gamma\Delta}{2}+\frac{1}{\pi}\left(\sum_{\mu=1}^{\mu_+}+\sum_{\mu=1}^{\mu_-}\right)\frac{1}{\mu}-\frac{1}{\pi\Delta}\sum_{\mu=\mu_++1}^{\infty}\frac{\left(\frac{1}{2}\frac{\Delta}{\mu}-\sin\psi_i\right)}{\left(\frac{\mu}{\Delta}\right)\left[\frac{\mu}{\Delta}+\sin\psi_i-\frac{1}{2}\frac{\Delta}{\mu}\right]}$$

$$-\frac{1}{\pi\Delta}\sum_{\mu=\mu_-+1}^{\infty}\frac{\left(\frac{1}{2}\frac{\Delta}{\mu}+\sin\psi_i\right)}{\left(\frac{\mu}{\Delta}\right)\left[\frac{\mu}{\Delta}-\sin\psi_i-\frac{1}{2}\frac{\Delta}{\mu}\right]} \tag{26}$$

In addition, we can obtain the simplified expressions for $\mathsf{N}_{2n}$ and $\mathsf{N}_{2n+1}$ in (24a, b) as

$$\mathrm{N}_{2n} \cong \frac{1}{n\pi} + \frac{1}{\pi}\sum_{m=1}^{n} \frac{(-1)^m 2^{2m}(n+m-1)!}{(2m)!\ (n-m)!} \frac{B_{2m}(\Delta \sin\psi_i)}{\Delta^{2m}}$$

$$-\frac{1}{\pi}\left(\sum_{\mu=-\mu_-}^{-1} - \sum_{\mu=0}^{\mu_+}\right)\sum_{m=1}^{n}\left[\frac{(-1)^m 2^{2m-1}(n+m-1)!}{(2m-1)!\ (n-m)!\Delta^{2m}}\right](\mu + \Delta\sin\psi_i)^{2m-1}$$

$$-\frac{(-1)^n}{\pi\Delta}\left\{\sum_{\mu=\mu_++1}^{\infty} \frac{\left(\frac{\mu}{2\Delta}\right)^{2n} + O\left(\left(\frac{\Delta}{\mu}\right)^2\right)}{\left(\frac{\mu}{\Delta}\right) + \sin\psi_i - \frac{1}{2}\left(\frac{\Delta}{\mu}\right) + O\left(\left(\frac{\Delta}{\mu}\right)^2\right)}\right.$$

$$\left. + \sum_{\mu=\mu_-+1}^{\infty} \frac{\left(\frac{\mu}{2\Delta}\right)^{2n} + O\left(\left(\frac{\Delta}{\mu}\right)^2\right)}{\left(\frac{\mu}{\Delta}\right) - \sin\psi_i - \frac{1}{2}\left(\frac{\Delta}{\mu}\right) + O\left(\left(\frac{\Delta}{\mu}\right)^2\right)}\right\} \qquad (27a)$$

$\forall n \ni n \in N$, and

$$\mathrm{N}_{2n+1} \cong \frac{2}{i\pi}\sum_{m=0}^{n} \frac{(-1)^m 2^{2m}(n+m)!}{(2m+1)!\ (n-m)!} \frac{B_{2m+1}(\Delta \sin\psi_i)}{\Delta^{2m+1}}$$

$$-\frac{1}{i\pi}\left(\sum_{\mu=-\mu_-}^{-1} - \sum_{\mu=0}^{\mu_+}\right)\sum_{m=0}^{n}\left[\frac{(-1)^m 2^{2m}(n+m)!}{(2m)!\ (n-m)!\Delta^{2m+1}}\right](\mu + \Delta\sin\psi_i)^{2m}$$

$$-\frac{(-1)^n}{i\pi\Delta}\left\{\sum_{\mu=\mu_++1}^{\infty} \frac{\left(\frac{\mu}{2\Delta}\right)^{2n} + O\left(\left(\frac{\Delta}{\mu}\right)^2\right)}{\left(\frac{\mu}{\Delta}\right) + \sin\psi_i - \frac{1}{2}\left(\frac{\Delta}{\mu}\right) + O\left(\left(\frac{\Delta}{\mu}\right)^2\right)}\right.$$

$$\left. - \sum_{\mu=\mu_-+1}^{\infty} \frac{\left(\frac{\mu}{2\Delta}\right)^{2n} + O\left(\left(\frac{\Delta}{\mu}\right)^2\right)}{\left(\frac{\mu}{\Delta}\right) - \sin\psi_i - \frac{1}{2}\left(\frac{\Delta}{\mu}\right) + O\left(\left(\frac{\Delta}{\mu}\right)^2\right)}\right\} \qquad (27b)$$

$\forall n \ni n \in Z_+$.

*3. 3. Special case with* $\mu_+ = \mu_- = 0$

If there is only one propagating mode, which corresponds to the physical problem when the scattering of wavelengths larger than the *'grating spacing',* i.e., $\frac{k_r d}{2\pi}(1 \pm \sin\psi_i) < 1$, then the *'Bessel series',* for $\phi_0 = \pi + \psi_i$, reduces to

$$\mathsf{J}_{2n} = \frac{2\cos 2n\phi_0}{k_r d\cos\phi_0} - \delta_{n0} \tag{28a}$$

$$\mathsf{J}_{2n+1} = -\frac{2i\sin(2n+1)\phi_0}{k_r d\cos\phi_0} \tag{28b}$$

$\forall n \ni n \in Z_+$.

*3. 3. 1. Approximations for* $\mathsf{N}_0$ , $\mathsf{N}_{2n}$ *and* $\mathsf{N}_{2n+1}$ *with the special case of* $\mu_+ = \mu_- = 0$ *in the limit of* $\Delta << 1$

Inserting $\mu_+ = \mu_- = 0$ into (26, 27a, and b), the expression for $\mathsf{N}_0$ in (26) reduces to

$$\mathsf{N}_0 \cong -\frac{2}{\pi}\ln\frac{\gamma\Delta}{2} - \frac{1}{\pi\Delta}\sum_{\mu=1}^{\infty}\left\{\frac{\left(1+2\sin^2\psi_i\right)-\frac{1}{2}\left(\frac{\Delta}{\mu}\right)^2}{\left(\frac{\mu}{\Delta}\right)^3\left[1-\left(1+\sin^2\psi_i\right)\left(\frac{\Delta}{\mu}\right)^2+\frac{1}{4}\left(\frac{\Delta}{\mu}\right)^4\right]}\right\} \tag{29}$$

The approximation for the *'Neumann series* $N_0$ *',* for $\phi_0 = \pi + \psi_i$, up to terms of the order $(k_r d)^2$ can be obtained from (9) as

$$\mathsf{N}_0 \cong -\frac{2}{\pi}\ln\frac{\gamma\Delta}{2} - \frac{\left(1+2\sin^2\psi_i\right)\Delta^2}{\pi}\sum_{\mu=1}^{\infty}\mu^{-3} \tag{30}$$

We have $\sum_{\mu=1}^{\infty}\mu^{-3} \approx 1.202$ in (30). In the same range, the *'Neumann series* $N_n$ *'* reduces to

$$\mathsf{N}_{2n} = \frac{1}{n\pi} + \frac{1}{\pi}\sum_{m=1}^{n}\frac{(-1)^m 2^{2m-1}(n+m-1)!}{(2m-1)!\,(n-m)!\,\Delta^{2m}}\left[\frac{B_{2m}(\Delta\sin\psi_i)}{m} + (\Delta\sin\psi_i)^{2m-1}\right] + \mathsf{F}_{2n} \tag{31a}$$

$\forall n \ni n \in N$, and

$$\mathsf{N}_{2n+1} = \frac{1}{i\pi}\sum_{m=0}^{n}\frac{(-1)^m 2^{2m}(n+m)!}{(2m)!\,(n-m)!\Delta^{2m+1}}\left[\frac{B_{2m+1}(\Delta\sin\psi_i)}{m+\frac{1}{2}}+(\Delta\sin\psi_i)^{2m}\right]+\mathsf{F}_{2n+1} \qquad (31b)$$

$\forall n \ni n \in Z_+$, where $\mathsf{F}$'s are given by

$$\mathsf{F}_{2n} \cong \frac{(-1)^{n+1}}{\pi\Delta}\sum_{\mu=1}^{\infty}\frac{1}{2^{2n-1}}\left(\frac{\Delta}{\mu}\right)^{2n+1} \qquad (32a)$$

$$\mathsf{F}_{2n+1} \cong i\frac{(-1)^{n+1}}{\pi\Delta}\sin\psi_i\sum_{\mu=1}^{\infty}\frac{1}{2^{2n}}\left(\frac{\Delta}{\mu}\right)^{2n+3} \qquad (32b)$$

*3. 3. 2. Approximations for the 'Schlömilch series',* $\mathsf{H}_n = \mathsf{J}_n + i\,\mathsf{N}_n$ *, with the special case of* $\mu_+ = \mu_- = 0$ *in the limit of* $\Delta << 1$

If $k_r d$ is small, that is to say if $\frac{k_r d}{2\pi}(1 \pm \sin\psi_i) < 1$, then there is only one discrete propagating mode. In this range, using the expansions for the *'Bessel and Neumann Series'* obtained in the previous sections, we can write the expansions for the *'Schlömilch Series'*, for $\phi_0 = \pi + \psi_i$, as

$$\mathsf{H}_0 \cong \frac{2}{k_r d\cos\phi_0} - \frac{2i}{\pi}\ln\frac{\gamma k_r d}{4\pi} - 1 - \frac{(k_r d)^2}{2\pi^3}\left(\frac{1}{2}+\sin^2\phi_0\right)1.202i \qquad (33a)$$

$$\mathsf{H}_1 \cong \frac{-2i\sin\phi_0}{k_r d\cos\phi_0} + \frac{2\sin\phi_0}{\pi} + \frac{(k_r d)^2\sin\phi_0}{4\pi^3}1.202 \qquad (33b)$$

$$\mathsf{H}_2 \cong \frac{4\pi}{3i(k_r d)^2} + \frac{2\cos 2\phi_0}{k_r d\cos\phi_0} + \frac{i}{\pi}(1-2\sin^2\phi_0) + \frac{(k_r d)^2}{(2\pi)^3}1.202i \qquad (33c)$$

$$\mathsf{H}_3 \cong -\frac{16\pi\sin\phi_0}{3(k_r d)^2} - i\frac{2\sin 3\phi_0}{k_r d\cos\phi_0} + \frac{2\sin\phi_0}{\pi}(1-\frac{4}{3}\sin^2\phi_0) - \frac{\sin\phi_0}{2}\frac{(k_r d)^4}{(2\pi)^5}\sum_{\mu=1}^{\infty}\mu^{-5} \qquad (33d)$$

$$\mathsf{H}_4 \cong \frac{2^5\pi^3}{15i(k_r d)^4} - i\frac{16\pi}{(k_r d)^2}(\frac{1}{6}-\sin^2\phi_0) + \frac{2\cos 4\phi_0}{k_r d\cos\phi_0} + \frac{i}{2\pi}(1-8\sin^2\phi_0+8\sin^4\phi_0) - i\frac{(k_r d)^4}{4(2\pi)^5}\sum_{\mu=1}^{\infty}\mu^{-5} \qquad (33e)$$

The leading terms of $\mathsf{H}$'s for large $\forall n \ni n \in N$ is given as

$$\mathsf{H}_{2n} \approx 2^{4n-1}\left[\frac{(-1)^n \pi^{2n-1} B_{2n}(0)}{(k_r d)^{2n}}\right]\frac{i}{n} \quad (34a)$$

$$\mathsf{H}_{2n+1} \approx 2^{4n+1}\left[\frac{(-1)^n \pi^{2n-1} B_{2n}(0)}{(k_r d)^{2n}}\right]\sin\phi_0 \quad (34b)$$

where $B_n(0)$ corresponds to *'Bernoulli Polynomial'*. From (33, and 34), we can determine the leading terms of the *'Schlömilch Series'* as

$$\mathsf{H}_0 \approx \frac{h_0}{k_r d} \quad \text{where} \quad h_0 \equiv 2\sec\phi_0 \quad (35a)$$

$$\mathsf{H}_1 \approx \frac{h_1}{k_r d} \quad \text{where} \quad h_1 \equiv -2i\tan\phi_0 \quad (35b)$$

$$\mathsf{H}_2 \approx \frac{h_2}{(k_r d)^2} \quad \text{where} \quad h_2 \equiv \frac{4\pi}{3i} \quad (35c)$$

$$\mathsf{H}_3 \approx \frac{h_3}{(k_r d)^2} \quad \text{where} \quad h_3 \equiv -\frac{16\pi\sin\phi_0}{3} \quad (35d)$$

$$\mathsf{H}_4 \approx \frac{h_4}{(k_r d)^4} \quad \text{where} \quad h_4 \equiv \frac{2^5\pi^3}{15i} \quad (35e)$$

$$\mathsf{H}_5 \approx \frac{h_5}{(k_r d)^4} \quad \text{where} \quad h_5 \equiv -\frac{2^8\pi^3\sin\phi_0}{15} \quad (35f)$$

The leading terms of $\mathsf{H}_n$ for large $n$ are given by

$$\mathsf{H}_{2n} \approx \frac{h_{2n}}{(k_r d)^{2n}} \quad (36a)$$

$$\mathsf{H}_{2n+1} \approx \frac{h_{2n+1}}{(k_r d)^{2n}} \quad (36b)$$

where $h_{2n}$'s and $h_{2n+1}$'s for large $n$ are given as

$$h_{2n} \to \frac{i}{n}(-1)^n 2^{4n-1}\pi^{2n-1}B_{2n}(0) \quad (37a)$$

and

$$h_{2n+1} \to (-1)^n 2^{4n+1}\pi^{2n-1}B_{2n}(0)\sin\phi_0 \equiv -4inh_{2n}\sin\phi_0 \quad (37b)$$

respectively. In the above expressions, $B_\xi$'s are the *'Bernoulli numbers',* and the relationship between *'Bernoulli polynomial'* and *'Bernoulli numbers'* is given as

$$B_{2\xi}(0) \equiv (-1)^{\xi-1} B_\xi \tag{38}$$

**4. Asymptotic expansions for the scattering coefficients of the infinite grating at oblique incidence in the limiting case of "$(a/d) << 1$"**

In order to find a solution for the set of equations given in (21 and 22), we have introduced an *'Ansatz'* for the scattering coefficients of the electric and magnetic fields of the infinite grating assuming $(k_r a) << 1$, and $\left(\frac{k_r a}{k_r d}\right) \equiv \xi < \frac{1}{2}$ as

$$A_{\pm(2n-1)} \cong A_{\pm(2n-1),0} (k_r a)^{2n} \tag{39a}$$

$$A^H_{\pm(2n-1)} \cong A^H_{\pm(2n-1),0} (k_r a)^{2n} \tag{39b}$$

$\forall n \ni n \in N$ for the odd coefficients, and

$$A_{\pm 2n} \cong A_{\pm 2n,0} (k_r a)^{2n+2} \tag{39c}$$

$$A^H_{\pm 2n} \cong A^H_{\pm 2n,0} (k_r a)^{2n+2} \tag{39d}$$

$\forall n \ni n \in Z_+$ for the even coefficients. . We have defined the overall effect of the multiple scattering terms when the wavelength is larger than the grating spacing, i.e., $(k_r d) << 1$, and $\left(\frac{k_r a}{k_r d}\right) \equiv \xi < \frac{1}{2}$ as

$$G_{\pm n} \equiv \sum_{m=-\infty}^{\infty} \mathrm{H}_{\pm n-m} A_m \tag{40a}$$

for the electric field coefficients, and

$$G^H_{\pm n} \equiv \sum_{m=-\infty}^{\infty} \mathrm{H}_{\pm n-m} A^H_m \tag{40b}$$

for the magnetic field coefficients. Introducing the Schlömilch Series into (14), we can write the overall effect of the multiple scattering terms when the wavelength is larger than the grating spacing, i.e., $(k_r d) << 1$, and $\left(\frac{k_r a}{k_r d}\right) \equiv \xi < \frac{1}{2}$ as

$$G_{\pm(2n-1),0} = \sum_{m=1}^{\infty} \left(\frac{a}{d}\right)^{2m} h_{\pm 2(m+n-1)} A_{\mp(2m-1),0} \tag{41a}$$

$$G^{H}_{\pm(2n-1),0} = \sum_{m=1}^{\infty}\left(\frac{a}{d}\right)^{2m} h_{\pm2(m+n-1)}A^{H}_{\mp(2m-1),0} \tag{41b}$$

$\forall n \ni n \in N$ for the odd coefficients;

$$G_{\pm2n,0} = \sum_{m=1}^{\infty}\left(\frac{a}{d}\right)^{2m}\left\{ h_{\pm2(m+n-1)}A_{\mp(2m-2),0} + h_{\pm(2m+2n-1)}A_{\mp(2m-1),0} \right\} \tag{41c}$$

$$G^{H}_{\pm2n,0} = \sum_{m=1}^{\infty}\left(\frac{a}{d}\right)^{2m}\left\{ h_{\pm2(m+n-1)}A^{H}_{\mp(2m-2),0} + h_{\pm(2m+2n-1)}A^{H}_{\mp(2m-1),0} \right\} \tag{41d}$$

$\forall n \ni n \in N$ for the even coefficients; and the special case for $n = 0$ is given by

$$G_{0,0} = \sum_{m=-1}^{1} h_m A_{-m,0} \tag{42a}$$

$$G^{H}_{0,0} = \sum_{m=-1}^{1} h_m A^{H}_{-m,0} \tag{42b}$$

Defining the wavelength independent parts of the scattering matrices from (13) as

$$\underline{\underline{S}}_{n} := \underline{\underline{S}}_{n,0}(k_r a)^{2n} \tag{43a}$$

$$\underline{\underline{S}}_{\pm n,0} := \frac{1}{D}\begin{pmatrix} s^{\varepsilon\mu}_{n} & s^{\xi}_{\pm n} \\ s^{\eta}_{\pm n} & s^{\mu\varepsilon}_{n} \end{pmatrix} \tag{43b}$$

$$\underline{\underline{S}}_{\pm(2n-1),0} \equiv \frac{1}{D}\begin{bmatrix} s^{\varepsilon\mu}_{2n-1,0} & s^{\xi}_{\pm(2n-1),0} \\ s^{\eta}_{\pm(2n-1),0} & s^{\mu\varepsilon}_{2n-1,0} \end{bmatrix} \tag{43c}$$

$\forall n \ni n \in N$ corresponding to the odd, and

$$\underline{\underline{S}}_{\pm2n,0} \equiv \frac{1}{D}\begin{bmatrix} s^{\varepsilon\mu}_{2n,0} & s^{\xi}_{\pm2n,0} \\ s^{\eta}_{\pm2n,0} & s^{\mu\varepsilon}_{2n,0} \end{bmatrix} \tag{43d}$$

$\forall n \ni n \in N$ corresponding to the even part. In terms of these definitions of (43), and Upon introducing (41) into (21 and 22) and employing the *'Kronocker delta $\delta_{nm}$ symbol'*, we have obtained the following set of equations for the approximations of the scattering coefficients as

$$\begin{bmatrix} A_{\pm(2n-1),0} \\ A^{H}_{\pm(2n-1),0} \end{bmatrix} = \underline{\underline{S}}_{\pm(2n-1),0}\begin{bmatrix} \delta_{n1}E^{i}_{\pm(2n-1),0} + \left(\frac{a}{d}\right)^{2(n-1)} G_{\pm(2n-1),0} \\ \left(\frac{a}{d}\right)^{2(n-1)} G^{H}_{\pm(2n-1),0} \end{bmatrix} \tag{44a}$$

$\forall n \ni n \in N$ corresponding to the odd scattering coefficients, and

$$\begin{bmatrix} A_{\pm 2n,0} \\ A^H_{\pm 2n,0} \end{bmatrix} = \underline{\underline{S}}_{\pm 2n,0} \begin{bmatrix} \delta_{n1} E^i_{\pm 2n,0} + \left(\frac{a}{d}\right)^{2(n-1)} G_{\pm 2n,0} \\ \left(\frac{a}{d}\right)^{2(n-1)} G^H_{\pm 2n,0} \end{bmatrix} \tag{44b}$$

$\forall n \ni n \in N$ corresponding to the even scattering coefficients. Splitting the matrices in (44) into two parts, we have

$$\begin{bmatrix} A_{\pm(2n-1),0} \\ A^H_{\pm(2n-1),0} \end{bmatrix} = \frac{\delta_{n1}}{D} \begin{bmatrix} s^{\varepsilon\mu}_{2n-1,0} \\ s^{\eta}_{\pm(2n-1),0} \end{bmatrix} E^i_{\pm(2n-1),0} + \left(\frac{a}{d}\right)^{2(n-1)} \underline{\underline{S}}_{\pm(2n-1),0} \begin{bmatrix} G_{\pm(2n-1),0} \\ G^H_{\pm(2n-1),0} \end{bmatrix} \tag{45a}$$

$\forall n \ni n \in N$ for the odd scattering coefficients, and

$$\begin{bmatrix} A_{\pm 2n,0} \\ A^H_{\pm 2n,0} \end{bmatrix} = \frac{\delta_{n1}}{D} \begin{bmatrix} s^{\varepsilon\mu}_{2n,0} \\ s^{\eta}_{\pm 2n,0} \end{bmatrix} E^i_{\pm 2n,0} + \left(\frac{a}{d}\right)^{2(n-1)} \underline{\underline{S}}_{\pm 2n,0} \begin{bmatrix} G_{\pm 2n,0} \\ G^H_{\pm 2n,0} \end{bmatrix} \tag{45b}$$

$\forall n \ni n \in N$ for even scattering coefficients, respectively. From (41), we have established the following terms as

$$\left(\frac{a}{d}\right)^{2(n-1)} G_{\pm(2n-1),0} = \sum_{m=1}^{\infty} \left(\frac{a}{d}\right)^{2(m+n-1)} h_{\pm 2(m+n-1)} A_{\mp(2m-1),0} \tag{46a}$$

for the multiple interactions corresponding to the scattering coefficients of the electric field,

$$\left(\frac{a}{d}\right)^{2(n-1)} G^H_{\pm(2n-1),0} = \sum_{m=1}^{\infty} \left(\frac{a}{d}\right)^{2(m+n-1)} h_{\pm 2(m+n-1)} A^H_{\mp(2m-1),0} \tag{46b}$$

for the multiple interactions corresponding to the scattering coefficients of the magnetic field, $\forall n \ni n \in N$ for the odd scattering coefficients; and

$$\left(\frac{a}{d}\right)^{2(n-1)} G_{\pm 2n,0} = \sum_{m=1}^{\infty} \left(\frac{a}{d}\right)^{2(m+n-1)} \left\{ h_{\pm 2(m+n-1)} A_{\mp(2m-2),0} + h_{\pm(2m+2n-1)} A_{\mp(2m-1),0} \right\} \tag{47a}$$

for the multiple interactions corresponding to the scattering coefficients of the electric field,

$$\left(\frac{a}{d}\right)^{2(n-1)} G^H_{\pm 2n,0} = \sum_{m=1}^{\infty} \left(\frac{a}{d}\right)^{2(m+n-1)} \left\{ h_{\pm 2(m+n-1)} A^H_{\mp(2m-2),0} + h_{\pm(2m+2n-1)} A^H_{\mp(2m-1),0} \right\} \tag{47b}$$

for the multiple interactions corresponding to the scattering coefficients of the magnetic field, $\forall n \ni n \in N$ for the even scattering coefficients. Inserting (46 and 47) into (45), we have finally obtained an infinite set of equations for the electric and magnetic scattering coefficients as

$$\begin{bmatrix} A_{\pm(2n-1),0} \\ A^H_{\pm(2n-1),0} \end{bmatrix} = \frac{\delta_{n1}}{D} \begin{bmatrix} s^{\varepsilon\mu}_{2n-1,0} \\ s^{\eta}_{\pm(2n-1),0} \end{bmatrix} E^i_{\pm(2n-1),0} + \sum_{m=1}^{\infty} \left(\frac{a}{d}\right)^{2(m+n-1)} h_{\pm 2(m+n-1)} \underline{\underline{S}}_{\pm(2n-1),0} \begin{bmatrix} A_{\mp(2m-1),0} \\ A^H_{\mp(2m-1),0} \end{bmatrix} \tag{48a}$$

$\forall n \ni n \in N$ for the odd scattering coefficients, and

$$\begin{bmatrix} A_{\pm 2n,0} \\ A^H_{\pm 2n,0} \end{bmatrix} = \frac{\delta_{n1}}{D} \begin{bmatrix} s^{\varepsilon\mu}_{2n,0} \\ s^{\eta}_{\pm 2n,0} \end{bmatrix} E^i_{\pm 2n,0} + \sum_{m=1}^{\infty} \left(\frac{a}{d}\right)^{2(m+n-1)} \underline{\underline{S}}_{\pm 2n,0} \left\{ h_{\pm 2(m+n-1)} \begin{bmatrix} A_{\mp 2(m-1),0} \\ A^H_{\mp 2(m-1),0} \end{bmatrix} + h_{\pm(2m+2n-1)} \begin{bmatrix} A_{\mp 2(m-1),0} \\ A^H_{\mp 2(m-1),0} \end{bmatrix} \right\} \tag{48b}$$

$\forall n \ni n \in N$ for the even scattering coefficients. In equations (48a, b), we have noticed that the scattering coefficients of the electric and magnetic fields appeared as coupled to each others. We have finally attempt to express the equations of (48) more compactly by defining a new vector such as $\underline{\varpi}_p$ where

$$\underline{\varpi}_p \equiv \begin{bmatrix} A_{p,0} \\ A^H_{p,0} \end{bmatrix} \quad \forall p \ni p \in Z \tag{49}$$

Inserting this definition of (49) into (48a, b), we have obtained

$$\underline{\varpi}_{\pm(2n-1),0} = \frac{\delta_{n1}}{D} \begin{bmatrix} s^{\varepsilon\mu}_{2n-1,0} \\ s^{\eta}_{\pm(2n-1),0} \end{bmatrix} E^i_{\pm(2n-1),0} + \left(\frac{a}{d}\right)^{2(n-1)} \underline{\underline{S}}_{\pm(2n-1),0} \sum_{m=1}^{\infty} \left(\frac{a}{d}\right)^{2m} h_{2(m+n-1)} \underline{\varpi}_{\mp(2m-1),0} \tag{50a}$$

$\forall n \ni n \in N$ for the odd scattering coefficients, and

$$\underline{\varpi}_{\pm 2n,0} = \frac{\delta_{n1}}{D} \begin{bmatrix} s^{\varepsilon\mu}_{2n,0} \\ s^{\eta}_{\pm 2n,0} \end{bmatrix} E^i_{\pm 2n,0} + \left(\frac{a}{d}\right)^{2(n-1)} \underline{\underline{S}}_{\pm 2n,0} \sum_{m=1}^{\infty} \left(\frac{a}{d}\right)^{2m} \left\{ h_{2(m+n-1)} \underline{\varpi}_{\mp 2(m-1),0} + h_{\pm(2m+2n-1)} \underline{\varpi}_{\mp(2m-1),0} \right\} \tag{50b}$$

$\forall n \ni n \in N$ for the even scattering coefficients.

## Conclusion

In this investigation, we have presented a rigorous derivation of the asymptotic equations associated with *'the scattering coefficients of an infinite grating of dielectric circular cylinders for obliquely incident vertically polarized plane electromagnetic waves'*. We have predicted the asymptotic behavior of the scattering coefficients by exploiting the *"Twersky's elementary function representations for Schlömilch series"* when the wavelength of the scattered wave is much larger than the distance between the constituent cylinders of the grating, and then used these predicted forms for the determination of an *'Ansatz'* which describes the behavior of the scattering coefficients as a function of $(a/d)$. Finally, we have acquired the asymptotic forms of the equations associated with the *'electric and magnetic scattering coefficients of the infinite grating at oblique incidence'*. Our results are the generalizations of those acquired by (Twersky 1962) for the non-oblique incidence case.

## Acknowledgments

We are indebted to deceased Professor Emeritus Victor Twersky of the Department of Mathematics of the University of Illinois at Chicago for the kind concern he spared to our work. The first author takes this opportunity to express his sincere thanks to Professor Dr Roger Henry Lang for suggesting the problem and many fruitful discussions.